\documentclass[oldversion,printer]{aa}

\usepackage{natbib}
\bibpunct{(}{)}{;}{a}{}{,}

\usepackage{epsfig,bm}
\usepackage{amssymb,amsmath}
\usepackage{subfigure}
\usepackage{txfonts}

\begin{document}

\title{Cosmological constraints on $f(R)$ gravity theories within the Palatini approach}
\author{M. Amarzguioui\inst{1} \and \O. Elgar\o y\inst{1,2} \and D. F. Mota\inst{1} \and
  T. Multam\"aki\inst{2,3}}
 \institute{Institute of Theoretical Astrophysics, University of Oslo, Box 1029, 0315 Oslo, Norway \and
NORDITA, Blegdamsvej 17, DK-2100, Copenhagen, Denmark \and
Department of Physics, University of Turku, FIN-20014, Turku, Finland}

\date{Received date / Accepted date}

\abstract{
We investigate $f(R)$ theories of gravity within the
Palatini approach and show how one can determine the
expansion history, $H(a)$, for an arbitrary choice of $f(R)$.
As an example, we consider cosmological constraints on such theories 
arising from the supernova type Ia, large scale structure formation and cosmic
microwave background observations. We find that best fit to the data is a 
non-null leading order correction to the Einstein gravity, but the current
data exhibits no significant preference over the concordance $\Lambda$CDM model.
Our results show that 
the often considered $1/R$ models are not compatible with the data.
The results demonstrate that the background expansion alone can act
as  a good discriminator between modified gravity models when 
multiple data sets are used.}


\authorrunning{M. Amarzguioui et al.}
\titlerunning{Cosmological constraints on $f(R)$ gravity theories within the Palatini approach}

\maketitle


\section{Introduction}

The combination of Einstein's General Relativity (GR) and ordinary matter,
as described by the standard model of particle physics
\citep{particlereview}, cannot explain the current cosmological
observations. Key observations confronting the matter only universe
are the luminosity-redshift relationship from observations of supernovae of 
type Ia (SNIa) \citep{Riess}, the matter power spectrum of large scale structure as inferred 
from galaxy redshift surveys like the Sloan Digital Sky Survey (SDSS) \citep{sloan} 
and the 2dF Galaxy Redshift Survey (2dFGRS) \citep{2dFGRS}, 
and the anisotropies in the Cosmic Microwave Background Radiation (CMBR) \citep{wmap}. 
In order to account for the results from all of these cosmological probes within
GR, two exotic components are required in the matter-energy
budget of the Universe. These two components are respectively
\emph{dark matter}, a collisionless and pressureless fluid, which
contributes about $25\%$ of the universe energy budget, and
a negative pressure fluid called \emph{dark energy}. 
Currently the dark energy component dominates the energy density
of the universe, causing accelerating expansion.

Despite the very good agreement between the so-called concordance model and
the astrophysical data, the nature of dark matter and dark energy is
one of the greatest mysteries of modern cosmology. 
In fact, none of the dark matter candidates from high energy physics 
beyond the standard model \citep{Bert04,Ellis98,Brook05,Amen99} have ever been observed.
One should bear in mind that the existence of
the dark sector is only inferred from the motion
of ordinary matter in a gravitational field. Hence one can ask whether 
the necessity of including dark matter and dark energy in the 
energy budget is a sign of our lack of understanding of
gravitational physics \citep{Lue03,Lue04,Lue02}. A natural alternative
to adding new exotic fluids is to modify gravitational
physics.
Several ways of modifying gravity in order to dispense with  
dark matter \citep{mond,mond1,mond2,mond3,mond4,mond5} or dark energy
\citep{Dval02,Deff00,Dval03,Arka02,Noji03b,Noji04,Noji03a,Noji05,Abda04,Free02,Meng03,Dolg03,Shao05,Gong04,Lima04,Ahme02,Bent02a,Bent02b} have been proposed, and 
they can with some degree of success account for the observations 
\citep{Lue03,Lue04,Lue02,Bent02a,Bent02b,Deff02,Carr04,Mult03a,Mult03b,Koiv04,Elga04,Amar04,Mult03c}.

In this article we investigate a family of alternative models to the dark
energy paradigm based on a generalization of the Einstein-Hilbert
Lagrangian. These models are called Nonlinear Theories of Gravity \citep{Magn93,Alle04}
or $f(R)$ theories \citep{Bron05,Cogn05,Nune04,Ezaw03,Barr00,Schm98,Ripp95,Barr83}, since the scalar curvature $R$ in the
Einstein-Hilbert Lagrangian is replaced by a general function
$f(R)$. The main motivation for this generalization is the fact that 
higher order terms in curvature invariants 
(such as $R^2$, $R^{\mu\nu}R_{\mu\nu}$,
$R^{\mu\nu\alpha\beta}R_{\mu\nu\alpha\beta}$, etc) have to be added to
the effective Lagrangian of the gravitational field when quantum
corrections are considered \citep{Buch92,Birr82,Vilk92,Gasp91}. Furthermore, 
there is no {\it a priori} reason to restrict oneself to the simple
Einstein-Hilbert action when a more general formulation is allowed.
Higher order terms in the gravitational action also have 
interesting consequences in cosmology, like natural inflationary behavior at early times \citep{Star80,Barr88,Barr91,La89}, and late time 
acceleration of the Universe \citep{Meng03,Meng03b,Carr04,Capo05}.  
Consequently, several authors have investigated whether 
such theories are indeed compatible with current cosmological
observations, big bang nucleosynthesis, and solar system constraints in
its weak field limit \citep{Clif05,Barr05,Quan90,Domi04,Soti05a,Soti05b,Olmo05a,Olmo05b,Capo02,Carl04,Barr02,Hwan01,Mena05,Alle04b,Alle04c,Alle05}.  
Most of these investigations have been model dependent and the
conclusions are somewhat contradictory \citep{Flan03,Voll03}.

When dealing with $f(R)$ theories of gravity, the choice of the
independent fields to vary in the action is a fundamental
issue. In the so-called Palatini approach one considers the metric and 
the connection to be independent of each other, and the resulting 
field equations are in general different from those one gets from
varying only the metric, the so-called metric approach. The two approaches
lead to the same equations only if $f(R)$ is linear in $R$. 
The correct choice of approach to derive the
field equations is still a hot topic of research. Initially, $f(R)$
theories were investigated in the metric approach. However, since this
method leads to fourth-order equations and the Palatini approach leads to
second-order equations, the latter is appealing because of its
simplicity. Moreover, the equations resulting from the metric approach
seem to have instability problems in many interesting cases
\citep{Dolg03,Chib03} from which the Palatini approach does not
suffer. However, recent work has cast doubts over these instabilities
\citep{Cemb05,Soti05a}. In the present paper we will concentrate on
the Palatini approach.  

The aim of this article is to use current cosmological data to
consider possible deviations from GR by combining a number of
different cosmological observations. The data used 
are the latest  Supernovae Ia gold set
\citep{Riess}, the CMBR shift parameter \citep{Bond97}, the baryon 
oscillation length scale \citep{Eise05} and the linear growth 
factor at the \mbox{2dFGRS} effective redshift \citep{Hawk02,Wang04}. 
As far as we are aware, this is 
the first time one uses all of the main cosmological data sets 
in order to constrain these models. 

The structure of the paper is as follows: In section~\ref{sec2} we deduce and
summarize the basic equations and properties of general $f(R)$
gravities in the Palatini approach. In section~\ref{sec3}, we investigate
observational constraints based on the evolution of the background of
the Universe. In particular, we consider fits to the SNIa and CMB 
shift parameter from the WMAP data. 
In section~\ref{sec4} we analyze the evolution of linear perturbations
and probe large scale structure formation in these models using the
linear growth factor derived from the 2dFGRS data. Finally,
section~\ref{sec5} contains a summary of our work and our conclusions. 

\section{General  $f(R)$ Gravity theories} \label{sec2}

The action that defines $f(R)$ gravity theories in the Palatini formalism is
the following: 

\begin{equation} \label{S}
S[f;g,\hat\Gamma,\psi_m] = -\frac{1}{2\kappa}\int d^4x \sqrt{-g}f(R) +
S_m[g_{\mu\nu},\psi_m] 
\end{equation} 
where  $\kappa=8\pi G$, $S_m[g_{\mu\nu},\psi_m]$ is the matter action
which depends only on the metric $g_{\mu\nu}$ and on the matter
fields $\psi_m$, $R\equiv R(g,\hat\Gamma) = 
g^{\mu\nu}R_{\mu\nu}(\hat\Gamma)$ is the generalized 
Ricci scalar and $R_{\mu\nu}(\hat\Gamma)$ is the Ricci tensor of the  
affine connection $\hat\Gamma$, which in the Palatini approach is
independent of the metric. 
The generalized Riemann tensor is given by \citep{Voll03b}:
\begin{equation} \label{Riemann} 
R^{\alpha}_{\mu\nu\beta}=\hat\Gamma^\alpha_{\mu\nu,\beta}-
\hat\Gamma^\alpha_{\mu\beta,\nu}+
\hat\Gamma^\lambda_{\mu\nu}\hat\Gamma^{\alpha}_{\beta\lambda}-
\hat\Gamma^\lambda_{\mu\beta}\hat\Gamma^{\alpha}_{\nu\lambda}.
\end{equation}
We define the Ricci tensor by contracting the first and the third
indices of the Riemann tensor. The field equations will be obtained
using the Palatini formalism i.e. we vary both with respect to the metric
and to the connection. 

Varying the above action with respect to
the metric, we obtain the generalized Einstein equations 
\begin{equation} \label{einstein}
f'(R)R_{\mu\nu}(\hat\Gamma)-\frac 12 f(R)g_{\mu\nu}=-\kappa T_{\mu\nu},
\end{equation}
where $f'(R)\equiv df/dR$ and $T_{\mu\nu}$ is the energy momentum
tensor
\begin{equation} \label{energymomemtum}
T_{\mu\nu} = -\frac{2}{\sqrt{-g}} \frac{\delta S_m}{\delta g^{\mu\nu}}.
\end{equation}
Varying with respect to the connection $\hat\Gamma$ and
contracting gives us the equation that determines the generalized
connection \citep{Voll03b}:
\begin{equation} \label{ei1}
\hat\nabla_{\alpha}[f'(R)\sqrt{-g}g^{\mu\nu}]=0,
\end{equation}
where $\hat\nabla$ is the covariant derivative with respect to the
affine connection $\hat\Gamma$. This equation implies that we can
write the affine connection as the Levi-Civita connection of a new
metric $h_{\mu\nu}=f'(R)g_{\mu\nu}$. The Levi-Civita connections of
the metrics $g_{\mu\nu}$ and $h_{\mu\nu}$ are then related by a
conformal transformation. This allows us to write the affine
connection as
\begin{equation} \label{twoconnetions}
\hat\Gamma^{\sigma}_{\mu\nu}=\Gamma^{\sigma}_{\mu\nu}+\frac{1}{2 f'} [2\delta^{\sigma}_{(\mu}\partial_{\nu)}f'-g^{\sigma\tau}g_{\mu\nu}\partial_{\tau}f'],
\end{equation}
where $\Gamma^{\sigma}_{\mu\nu}$ is the Levi-Civita connection of the metric $g_{\mu\nu}$.

The generalized Ricci tensor can now be written as
\begin{equation} \label{rmunu}
R_{\mu\nu}=R_{\mu\nu}(g)-\frac 32 \frac{\nabla_\mu f' \nabla_\nu f'}{f'^2}
+\frac{\nabla_\mu\nabla_\nu f'}{f'}+\frac 12 g_{\mu\nu}\frac{
\nabla^\mu\nabla_\mu f'}{f'}.
\end{equation}
Here $R_{\mu\nu}(g)$ is the Ricci tensor associated with $g_{\mu\nu}$,
 and $\nabla_\mu$ the covariant derivative associated with the
 Levi-Civita connection of the metric.

Since we are interested in cosmological solutions we consider the 
spatially flat FRW metric,
\begin{equation} \label{FRW}
ds^2=-dt^2+a(t)^2\delta_{ij}dx^idx^j.
\end{equation}
and the perfect fluid energy-momentum tensor $T_\mu^{\nu}=\textrm{diag}(-\rho,p,p,p)$.

Taking the trace of Eq. (\ref{einstein}), gives
\begin{equation} \label{contr}
R f'(R)-2f(R)=-\kappa T,
\end{equation}
where $T=g^{\mu\nu}T_{\mu\nu}=-(\rho-3p)$.
Eq. (\ref{contr}) can in certain special cases be solved 
explicitly for $R=R(T)$, but this is not possible in general.



The generalized Friedmann equation can be derived straightforwardly using the generalized
Ricci tensor, (\ref{rmunu}):
\begin{equation} \label{friedmann}
\left(H+\frac 12 \frac{\dot{f'}}{f'}\right)^2=\frac 16 \frac{\kappa (\rho+3p)}{f'}-\frac 16
\frac{f}{f'},
\end{equation}
which agrees with the result found in \citet{Alle04b,Alle04c} (note the
sign difference in the definition of the Einstein  equation)\footnote{
This differs from the results of \citet{Wang04b} by $3H\dot{f'}/f'$. 
}.

If the equation of state of the fluid, $p=p(\rho)$, is known,
one can use the continuity equation of the fluid together with Eq.~(\ref{contr})
to express $\dot{R}$ as
\begin{equation} \label{dotR}
\dot{R}=-3H\frac{\big(1-3p'(\rho)\big)\big(\rho+p(\rho)\big)}{R f''(R)-f'(R)}.
\end{equation}
Using the Friedmann equation it is easy to see that now $H=H(\rho,R)$, which together 
with Eq. (\ref{contr}) forms an algebraic set of equations from which one can
in principle determine $H=H(R)$ for any fluid.

In particular, in the case of a constant equation of state,
$p=w\rho$, we get 
$\kappa\rho=(Rf'-2f)/(1-3w)$ and hence the generalized Friedmann equation
can be written as

\begin{equation} \label{HRform}
H^2=
\frac{1}{6(1-3w)f'}
\frac{(1+3w)Rf'-3(1+w)f}
{\left(1-\frac 32 (1+w)\frac{f''(Rf'-2f)}{f'(Rf''-f')}\right)^2}.
\end{equation}
Similarly, from Eq. (\ref{contr}), we can write $a=a(R)$:
\begin{equation} \label{aRform}
a=\left(\frac{1}{\kappa\rho_0(1-3w)}\left(Rf'-2f\right)\right)^{-\frac{1}{3(1+w)}},
\end{equation}
where $\rho_0=\rho(a_0)$ and we have chosen $a_0=1$.
Using these two equations, (\ref{HRform}) and (\ref{aRform}),
it is now straightforward to determine the expansion history,
$H(a)$, for any $f(R)$.

In the special case $w=0$, applicable when the universe is matter dominated,
these equations reduce to
\begin{eqnarray} \label{HRform2}
H^2 & = & 
\frac{1}{6 f'}
\frac{Rf'-3f}
{\left(1-\frac 32 \frac{f''(Rf'-2f)}{f'(Rf''-f')}\right)^2},\\
a & = &  \left(\frac{1}{\kappa\rho_0}\left(Rf'-2f\right)\right)^{-\frac{1}{3}}.
\label{aform2}
\end{eqnarray}

\subsection{The leading correction to Einstein gravity}


In order to investigate to what extent observations allow deviations from
general relativity (where $f(R)=R$), we consider
the following gravity Lagrangian:
\begin{equation} \label{ans}
f(R)=R\left(1+\alpha\left(-\frac{R}{H_0^2}\right)^{\beta-1}\right),
\end{equation}
where $\alpha,\ \beta$ are dimensionless parameters (note that in out notation
$R$ is negative). 
Specializing to a matter dominated universe ($w=0$), we get from Eq. (\ref{contr})
the relation between the curvature scalar and matter density:
\begin{equation} \label{kaprho}
\kappa\rho_m=-R\left(1+\alpha(2-\beta)\left(-\frac{R}{H_0^2}\right)^{\beta-1}
\right).
\end{equation}
We wish to recover standard behaviour at early times. This implies
that the correction term in the Lagrangian must vanish for large $|R|$,
and hence we must demand that $\beta<1$. Furthermore, we must demand
that the right hand sides of equations \eqref{HRform2} and
\eqref{kaprho} always remain positive, which restricts the parameter space further.

Defining $\Omega_m\equiv \kappa\rho_m^0/(3 H_0^2)$, and choosing
units so that $H_0=1$, we can solve for $R_0$ from Eq. 
(\ref{kaprho}). Consistency then requires that substituting the obtained
value of $R_0$ into Eq. (\ref{HRform}) must give $H_0=1$. Hence,
given $\alpha$ and  $\beta$, $\Omega_m$ is fixed.

As an example, consider the case $\beta=0$, which corresponds to
the $\Lambda$CDM model. From Eq. ({\ref{HRform}), we have 
$H_0^2=(3\alpha H_0^2-2R_0)/6$ and from Eq. (\ref{kaprho}), 
$3\Omega_m H_0^2=2\alpha H_0^2-R_0$, and hence $\Omega_m=1+\alpha/6$
is fixed.

\section{Observational constraints from the Background Evolution} \label{sec3}

Armed with the modified Friedmann equation, we can now consider the
constraints arising from cosmological observations. In this section we
will consider quantities related to the background expansion of the Universe: 
the SNIa luminosity distance-redshift relationship, the CMBR shift parameter 
and the baryon oscillation length scale.

\subsection{CMBR Shift Parameter}
\begin{figure}
\resizebox{\hsize}{!}{\includegraphics{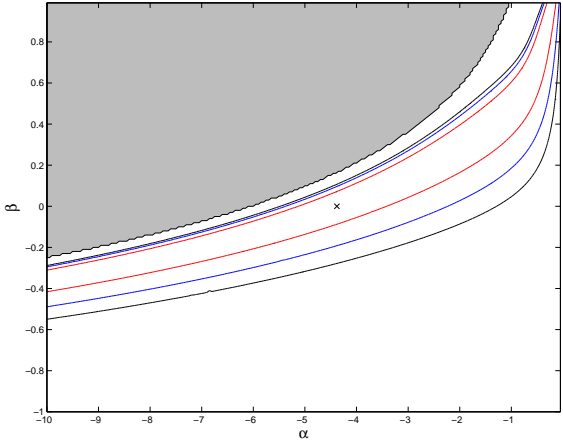}}
\caption{The $68,\, 95$ and $99\%$ confidence contours arising
from fitting the CMBR  shift parameter. The parameter values
corresponding to the concordance $\Lambda$CDM model ($\Omega_m=0.27,\
\Omega_{\Lambda}=0.73$ or $\alpha=-4.38,\ \beta=0$) are marked with a
cross. The gray area represents a section of the parameter space that
is not allowed.}
\label{figshift}
\end{figure}

The CMBR shift parameter \citep{Bond97,Melc02,Odma02} in a spatially 
flat universe is given by 
\begin{equation} \label{shiftdef}
{\cal R} = \sqrt{\Omega_m H_0^2}\int_0^{z_{dec}}\frac{d\tilde{z}}{H(\tilde{z})},
\end{equation}
where $z_{dec}$ is the redshift at decoupling.
The WMAP team \citep{wmap} quotes $z_{dec}=1088^{+1}_{-2}$ and
${\cal R}=1.716\pm 0.062$. In writing the shift parameter in this form we have implicitly assumed that photons follow geodesics determined by the Levi-Civita connection. In \citet{Koiv05a} this is shown to be the case if there is no torsion present. Furthermore, in order to use the shift parameter, the evolution of the universe needs to be standard up to very late times so 
that at decoupling we recover standard matter dominated behaviour.
Hence, we restrict our analysis to $\beta<1$.

Since we do not have an explicit expression for the
Hubble parameter in terms of the redshift, it is useful to rewrite
the shift parameter in terms of the curvature scalar, $R$:
\begin{eqnarray} \label{shift}
{\cal R} & = & \sqrt{\Omega_m H_0^2}\int_{0}^{z_{dec}}\frac{dz}{H(z)}\nonumber\\
& = & \sqrt{\Omega_m H_0^2}\int_{R_{dec}}^{R_0}\frac{a'(R)}{a(R)^2}
\frac{dR}{H(R)}\\
& = & \frac{1}{3^{4/3}}\left(\Omega_m H_0^2\right)^{1/6}\int_{R_0}^{R_{dec}}
\frac{Rf''-f'}{\left(Rf'-2f\right)^{2/3}}\frac{dR}{H(R)}\nonumber.
\end{eqnarray}

The constraints arising from the CMBR shift parameter can be seen in figure \ref{figshift}.

\subsection{SNIa constraints}

In order to incorporate measurements from SNIa, it is useful to rewrite the expression for the
luminosity distance as
\begin{eqnarray} \label{lumdist}
D_L(z) & = & (1+z)\int_{0}^z\frac{d\tilde{z}}{H(\tilde{z})}\nonumber\\
& = & \sqrt{\Omega_m H_0^2}\frac{1}{a(R)}\int_{R}^{R_0}\frac{a'(R)}{a(R)^2}\\
& = & \frac{1}{3}\sqrt{\Omega_m H_0^2}\left(Rf'-2f\right)^{1/3} \times \nonumber\\&&\int_{R_0}^{R}
\frac{Rf''-f'}{\left(Rf'-2f\right)^{2/3}}\frac{dR}{H(R)}\nonumber.
\end{eqnarray}

For the supernova data, we use the ``Gold data set'' from
\citet{Riess}. The contour plots showing the constraints on the
parameters from the supernovae can be seen in figure~\ref{figsn}. To
get these plots we marginalized over the Hubble parameter $h$.
\begin{figure}
\resizebox{\hsize}{!}{\includegraphics{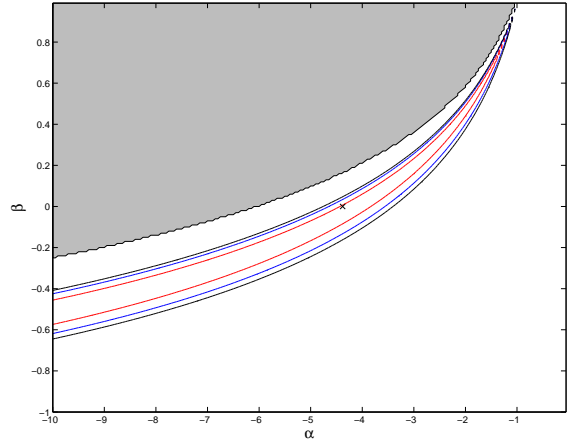}}
\caption{The $68,\, 95$ and $99\%$ confidence contours arising
from fitting the SN Ia data. The parameter values
corresponding to the concordance $\Lambda$CDM model ($\Omega_m=0.27,\
\Omega_{\Lambda}=0.73$ or $\alpha=-4.38,\ \beta=0$) are marked with a
cross. The gray area represents a section of the parameter space that
is not allowed.} 
\label{figsn}
\end{figure}

With the added information from the CMBR in the form of the shift parameter,
the situation improves as one can see from from figure~\ref{figsnshift}.
Still quite a large degeneracy persists on the $99\%$ level, but 
on the $68\%$ level, the model is quite well constrained and centered
around the concordance $\Lambda$CDM model.
\begin{figure}
\resizebox{\hsize}{!}{\includegraphics{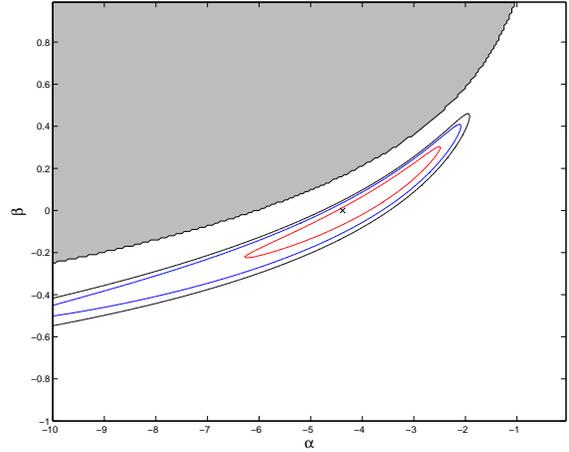}}
\caption{The combined $68,\, 95$ and $99\%$ confidence contours arising
from fitting the SN Ia and the CMBR shift parameter data. The parameter values
corresponding to the concordance $\Lambda$CDM model ($\Omega_m=0.27,\
\Omega_{\Lambda}=0.73$ or $\alpha=-4.38,\ \beta=0$) are marked with a
cross. The gray area represents a section of the parameter space that
is not allowed.}
\label{figsnshift}
\end{figure}

\subsection{Baryon Oscillations}

The baryon oscillations in the galaxy power spectrum are imprints from
acoustic oscillations prior to recombination, which are also responsible for the
the acoustic peaks seen in the CMBR temperature power spectrum. 
The physical length scale associated with the oscillations is set by the
sound horizon at recombination, which can be estimated from the
CMBR data \citep{wmap}. Measuring the apparent size of the
oscillations in a galaxy survey allows one to measure
the angular diameter distance at the survey redshift.
Together with the angular size of the CMB sound horizon, the
baryon oscillation size is a powerful probe of the properties and
evolution of the universe.

The imprint of the primordial baryon-photon acoustic oscillations 
in the matter power spectrum provides us therefore with a `standard ruler'
via the dimensionless quantity $A$ \citep{Lind03,Lind05,Hu95,Eise97,Eise04}:
\begin{equation}
A= \sqrt{\Omega_m}E(z_1)^{-1/3}\left[ \frac{1}{z_1}\int_{0}^{z_1}\frac{dz}{E(z)}\right]^{2/3},
\end{equation}
where $E(z)=H(z)/H_0$.

Recently the acoustic signature associated with the baryonic oscillations 
has been identified at low redshifts in the
distributions of Luminous Red Galaxies in the Sloan Digital Sky Survey
\citep{Eise05}, with a value of 
\begin{equation} \label{boe}
A=D_v(z=0.35)\frac{\sqrt{\Omega_m H_0^2}}{0.35c}=0.469\pm0.017,
\end{equation}
where 
\begin{equation} \label{distance}
D_v(z)=\left[D_M(z)^2\frac{cz}{H(z)}\right]^{1/3},
\end{equation}
and $D_M(z)$ is the comoving angular diameter distance.
For instance, in the case of $\Lambda$CDM with $\Omega_m=0.3$, $\Omega_{\Lambda}=0.7$ and $h=0.7$ we have $D_v(z=0.35)=1334$Mpc. 

Using the observed baryon oscillation length scale, one can hence
constrain the cosmological model.
The confidence contours for the modified gravity model we are
concerned with in this paper,
are shown in figure \ref{figbaryon}.
\begin{figure}
\centering
\resizebox{\hsize}{!}{\includegraphics{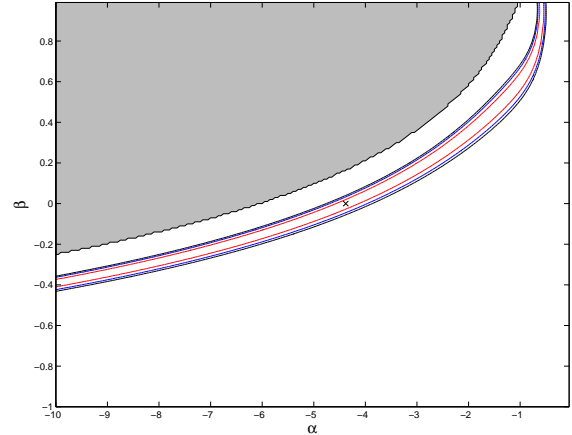}}
\caption{The $68,\, 95$ and $99\%$ confidence contours arising
from fitting the length scale associated with the baryon oscillations. The parameter values
corresponding to the concordance $\Lambda$CDM model ($\Omega_m=0.27,\
\Omega_{\Lambda}=0.73$ or $\alpha=-4.38,\ \beta=0$) are marked with a
cross. The gray area represents a section of the parameter space that
is not allowed.}
\label{figbaryon}
\end{figure}

Once again there is a large degeneracy in the $\alpha$ vs $\beta$ plane,
similar to the case when we fit SNIa (figure \ref{figsn}) and the CMBR
shift parameter (figure \ref{figshift}). Such degeneracies are
strongly restricted, however, when one combines all of the data sets 
in one single plot (see figure~\ref{figbosnsf}). 
\begin{figure}
\resizebox{\hsize}{!}{\includegraphics{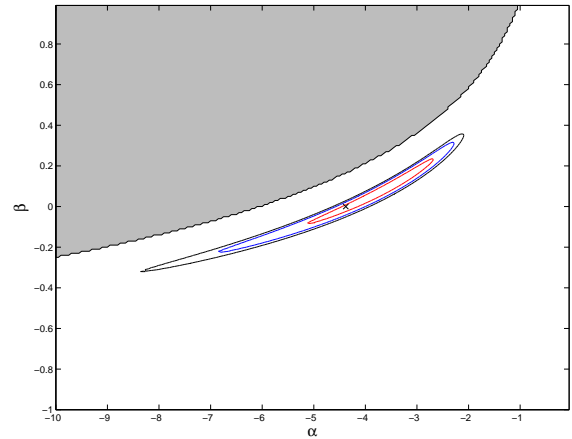}}
\caption{The $68,\, 95$ and $99\%$ confidence contours arising
from fitting the combined data from the SNIa, CMBR shift parameter and
the length scale associated to the baryon oscillations. The parameter
values corresponding to the concordance $\Lambda$CDM model
($\Omega_m=0.27,\ \Omega_{\Lambda}=0.73$ or $\alpha=-4.38,\ \beta=0$) are
marked with a cross. The gray area represents a section of the
parameter space that is not allowed.}
\label{figbosnsf}
\end{figure}

Using the baryon oscillations length scales and combining it with the
SNIa and the CMBR shift parameter data hence imposes strong constraints on
possible deviations from Einstein's General Relativity. This
also demonstrates how combining the current data, one can efficiently 
study and constrain cosmological models from the background expansion only.

The best fit model to the three data sets is $\alpha=-3.6$ and
$\beta=0.09$. This is slightly different from the $\Lambda$CDM model,
but the $\Lambda$CDM model is well within the $68\%$ confidence contour.

\section{Large Scale Structure: Formalism and Constraints} \label{sec4}

So far we have only considered observables related to the 
background evolution. In order to gain further information, 
it is useful to go beyond these `zeroth order' tests and consider 
perturbations within $f(R)$ models. Knowledge on the evolution
of perturbations allows one to confront models with large
scale structure observations from galaxy surveys.

Several authors have investigated cosmological
perturbations in generalized gravity theories in the metric approach
\citep{hwan91a,hwan91b}. However within the Palatini formalism, only very
recently the first steps were made \citep{Koiv05b}. 
Here we will follow 
a spherical collapse formalism \citep{Lue03,Mult03a}
where by requiring that a general gravity theory respects the 
Jebsen-Birkhoff theorem\footnote{What is commonly known as Birkhoff's theorem was first probed by the Norwegian J. T. Jebsen in 1921, two years prior to Birkhoff's work. See physics/0508163 for more historical details.} one derives the modified gravitational force law necessary to describe the evolution of the density perturbations at astrophysical scales. 

\subsection{Spherical collapse in $f(R)$ theories}

Although the spherical collapse model has been set and 
used for a long time \citep{Peeb93,Padm93} and in many different 
contexts \citep{Lue03,Mota03a,Mota03b,Mota04,Clif04b}, 
it has not been applied previously to Nonlinear Gravity Theories.
In \citet{Lue03}, one starts by assuming a
generalization of the Jebsen-Birkhoff theorem: for
any test particle outside a spherically symmetric matter source, the
metric observed by that test particle is equivalent to that of a point
source of the same mass located at the center of the sphere.  With
this one assumption, one can deduce the
Schwarzschild-like metric of the new hypothetical gravity theory.
Armed with the Schwarzschild-like metric one can then investigate the
evolution of spherical matter overdensities and compare it to the
latest large scale structure data. The idea is the following: Consider
a uniform sphere of dust.  Imagine that the evolution inside the
sphere is exactly cosmological, while outside the sphere is empty
space, whose metric (given the Jebsen-Birkhoff theorem) is 
Schwarzschild-like (as
defined by the metric equation~(2.2) in \cite{Lue03}). The mass of the
matter source (as determined by the form of the metric at short
distances) is unchanged throughout its time-evolution. The surface of
the spherical mass therefore charts out the metric through all of
space as the sphere expands with time, as long as we demand that the
cosmological metric just inside the surface of the sphere smoothly
matches the Schwarzschild solution just outside. In order to see how
the metric depends on the mass of the central source, we just take a
sphere of dust of a different initial size, and watch its surface
chart out a new metric. The procedure for determining the metric from
the cosmological evolution is described in \citet{Lue03} and we refer the reader to this article for further information and details.

An open question is the validity of the Jebsen-Birkhoff theorem
in $f(R)$ theories.
Although one can not
explicitly show that the only possible solution to the field
equations, when a spherically symmetric ansatz is inserted into them,
is the Schwarzschild metric, there is strong evidence that this is 
indeed the case. In fact that was shown for the case of
$f(R)=R+R^2$, where
$R^2=R^{\beta}\,_{\alpha}\,^{\mu\nu}R^{\alpha}\,_{\beta\mu\nu}$
\citep{Rama79}. And it was generalized later for any type of 
invariant of the form $R^2$, even in the case of a non-null torsion
\citep{Nevi79}. The authors also claimed that similar results
would most probably be valid even for the case of higher order
curvature invariants, such as $R^3$, $R^4$, etc. However, there is no
mathematical proof of this as of yet, even though there are several
studies and proofs for other complex cases such as multidimensional
gravities, Einstein-Yang-Mills systems, and conformally transformed
metrics \citep{Brod92,Bron94}.

 
\subsubsection{Growth of perturbations}

We want to follow the evolution of a lump of matter throughout the history
of our Universe. We start by considering a Universe with a 
Schwarzschild-like metric.
\begin{equation} \label{line-static}
     ds^2 = g_{00}(t,r)dT^2 - g_{rr}(t,r)dr^2 - r^2d\Omega\ ,
\end{equation}
where the metric components are uniquely determined from a given 
$a(t)$ \citep{Lue03}:
\begin{eqnarray} \label{metric-rr}
     g_{00} &=& E^2[1 - (\frac{r}{a})^2\dot{a}^2]
     \\ \nonumber
     g_{rr}^{-1} &=& 1 - (\frac{r}{a})^2\dot{a}^2\ ,
\end{eqnarray}
where $E=\sqrt{g_{00}g_{rr}}$ is a ${\rm constant}$.  

In order to investigate the evolution of density perturbations in this
scenario we use the spherical collapse model \citep{Peeb93}.
Consider a  top-hat overdensity $\delta(t)$ of a spherical distribution of
dust with mass $M$ and radius $r$ defined by
\begin{equation} \label{delta1}
	1+\delta = \frac{M}{\frac{4\pi}{3}\rho r^3}\ ,
\end{equation}
where ${\rho}(t)$ is the background matter density. Using the geodesic 
equation as expressed by differentiating Eq.~(\ref{metric-rr}) with
respect to $t$, we get
\begin{equation} \label{geodesic}
     \ddot{r} = -\frac{1}{2}\frac{d}{dr} g_{rr}^{-1}
     = rH_0^2\left[g(x)-\frac{3}{2}xg'(x)\right]\ .
\end{equation}
where $g(x)$ is defined via a generalized Friedman equation
\begin{equation} \label{mod-fried}
     g\left( x\right) = \frac{H^2}{H_0^2},
\end{equation}
and where we have expressed $\rho$ in terms of a dimensionless
quantity $x$ defined as 
\begin{equation} \label{xdef}
    {x} \equiv \frac{8\pi G{\rho}}{3H_0^2}=\Omega_m
{a}^{-3}
\end{equation}
Using the constraint from the Jebsen-Birkhoff theorem, 
one can calculate the evolution
of an overdensity by following the geodesic of a spherical mass,
without the need to consider what is happening outside the spherical
mass itself.  This is only possible if spherically symmetric
configurations respect the metric equations~(\ref{metric-rr})
\citep{Lue03}. Differentiating equation (\ref{delta1}) twice with respect to
time, and using  equation (\ref{geodesic}), one obtains a new 
equation for $\delta(t)$ 
\begin{eqnarray} \label{mg}
     \ddot{\delta}+2{H}\dot{\delta}
     -\frac{4}{3}\frac{1}{1+\delta}\dot{\delta}^2
     = \qquad \qquad \qquad \qquad  \qquad \nonumber\\3(1+\delta)H_0^2\left[
       \frac{3}{2}{x}(1+\delta)g'({x}(1+\delta))
       - g({x}(1+\delta))\right] \nonumber \\
-  3(1+\delta)H_0^2\left[\frac{3}{2}{x}g'({x}) - g({x})\right]\ .
\end{eqnarray} 
At linear order in perturbation theory, equation~(\ref{mg}) gives:
\begin{equation} \label{LPT}
     \ddot{\delta}+2{H}\dot{\delta}
     = 4\pi G{\rho}\delta\left[
       g'({x}) + 3{x}g''({x})\right]\ .
\end{equation}
In our case both $H$ and $\rho$ are expressed as functions of the
scalar curvature $R$ and not the time explicitly. We therefore
need to rewrite \eqref{LPT} in terms of derivatives of $R$. The
equation to be solved will then be of the form
\begin{eqnarray} \label{LPT_R}
     \frac{d^2\delta}{dR^2}+A(R)\frac{d\delta}{dR}=B(R)\delta\,.
\end{eqnarray}
where $A(R)$ and $B(R)$ are two rather unattractive functions of the
scalar curvature.
Using Eq. (\ref{LPT_R}), one can solve for the evolution of the 
linear growth factor and compare to observations.


\subsubsection{Constraints}

The large scale structure information we choose to use here 
is the linear growth rate  $F(z_{2dF})=0.51\pm0.11$ measured by the 2dFGRS \citep{Verd01,Knop03,Hawk02}, where $F \equiv d \ln D/d
lna$. We compare the theoretical value we get for the linear growth
rate for our model with the value measured by the 2dFGRS at its
effective redshift, $z_{2dF}=0.15$. The constraints arising from the
linear growth rate are plotted in figure~\ref{figlg}. 
\begin{figure}
\resizebox{\hsize}{!}{\includegraphics{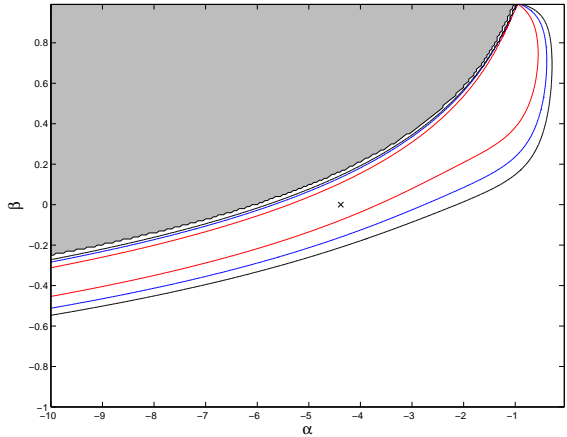}}
\caption{The $68,\, 95$ and $99\%$ confidence contours arising
from fitting the linear growth factor using the spherical collapse approach. The parameter values
corresponding to the concordance $\Lambda$CDM model ($\Omega_m=0.27,\
\Omega_{\Lambda}=0.73$ or $\alpha=-4.38,\ \beta=0$) are marked with a
cross. The gray area represents a section of the
parameter space that is not allowed.}
\label{figlg}
\end{figure}
Combining this with all the other constraints leads to the
confidence contours shown in Fig. \ref{figall}.

\begin{figure}
\resizebox{\hsize}{!}{\includegraphics{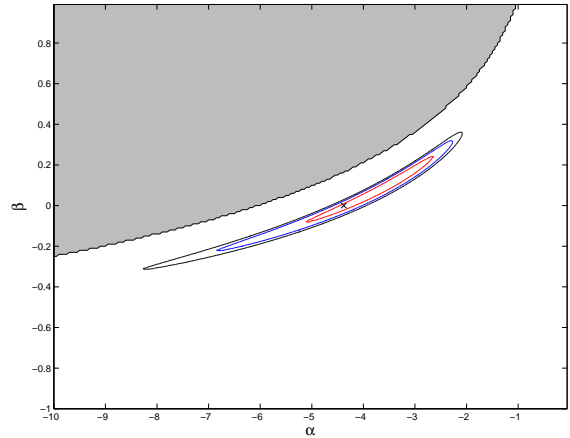}}
\caption{The $68,\, 95$ and $98\%$ confidence contours arising
from fitting to the Combined SNIa, shift parameter, baryon
oscillations and the linear growth data sets. The parameter values 
corresponding to the concordance $\Lambda$CDM model ($\Omega_m=0.27,\
\Omega_{\Lambda}=0.73$ or $\alpha=-4.38,\ \beta=0$) are marked with a
cross. The gray area represents a section of the
parameter space that is not allowed.}
\label{figall}
\end{figure}

\section{Conclusions} \label{sec5}

We have investigated observational constraints on $f(R)$ theories 
within the Palatini formalism. 
In order to relate these theories to observations, we have
shown how one can determine the expansion history for a given $f(R)$.
In particular, in a matter dominated universe, determining $H(a)$
is straightforward as expressed by Eqs (\ref{HRform2}) and (\ref{aform2}).

In particular we have investigated the possible form of the 
leading correction to standard GR, 
parameterizing the Lagrangian for gravity as 
${\mathcal{L}}_G=R-\alpha (-R/H_0^2)^{\beta}$, and have used a 
combination of data sets to determine the allowed ranges of $\alpha$ and 
$\beta$. This is by no means an exhaustive study. Other
interesting forms of $f(R)$ that are definitely worth studying include
$f(R)=\ln(R)$ and perhaps especially $f(R)=R-c_1/R+c_2 R^3$
\citep{Soti05b}, but our main purpose here has been to set up the
formalism and demonstrate the effectiveness of combining the current
data sets. 

Using a combination of data sets that probe the background
evolution, 
we have found that the current data efficiently constrains the
allowed parameter space of the leading correction to GR in the 
Palatini approach.
The best fit models to the individual data
sets are $(\alpha,\beta)=(-10.0,-0.51)$ for the SNIa,
$(\alpha,\beta)=(-8.4,-0.27)$ for the CMBR shift parameter and
$(\alpha,\beta)=(-1.1,0.57)$ for the baryon oscillations.
The best fit to the combination of these data sets is
$(\alpha,\beta)=(-3.6,0.09)$, but the $\Lambda$CDM concordance model
is well within the $1\sigma$ contour. Note, however, that the commonly
considered $1/R$ model is strongly disfavoured by the data.

In order to bring in additional information from the current 
galaxy surveys, we have also considered the growth of structures
in these models of modified gravity.
By assuming that the new gravitational physics obeys
a limited version of the Jebsen-Birkhoff theorem, 
we can describe the evolution of
overdensities in $f(R)$ gravity theories at 
sub-horizon scales. We find the best fit model to the linear
growth factor alone to be $(\alpha,\beta)=(-4.25,0.05)$ but
the allowed parameter range is degenerate and does not improve
constraints derived from the background evolution.
In order to fully utilize the information available from the
galaxy survey in form of the large scale matter power spectrum,
a more detailed analysis is needed along the lines presented in
\citep{Koiv05b}.

In summary, modified gravities provide us with an interesting alternative to the cosmological concordance model with a dominant 
dark energy component. Modern cosmological data can efficiently
constrain such models. These data indicate that currently there is no
compelling  evidence for non-standard gravity.

\begin{acknowledgements}
We would like to thank T. Koivisto for useful discussion.
MA, {\O}E and DFM acknowledge support from the Research Council of Norway 
through project numbers 159637/V30 and 162830/V00. TM is supported by
the Academy of Finland through project number 108658.
\end{acknowledgements}



\begin{thebibliography}{113}
\expandafter\ifx\csname natexlab\endcsname\relax\def\natexlab#1{#1}\fi

\bibitem[{Abdalla {et~al.}(2005)Abdalla, Nojiri, \& Odintsov}]{Abda04}
Abdalla, M.~C.~B., Nojiri, S., \& Odintsov, S.~D. 2005, Class. Quant. Grav.,
  22, L35

\bibitem[{Ahmed {et~al.}(2004)Ahmed, Dodelson, Greene, \& Sorkin}]{Ahme02}
Ahmed, M., Dodelson, S., Greene, P.~B., \& Sorkin, R. 2004, Phys. Rev., D69,
  103523

\bibitem[{Allemandi {et~al.}(2004{\natexlab{a}})Allemandi, Borowiec, \&
  Francaviglia}]{Alle04b}
Allemandi, G., Borowiec, A., \& Francaviglia, M. 2004{\natexlab{a}}, Phys.
  Rev., D70, 043524

\bibitem[{Allemandi {et~al.}(2004{\natexlab{b}})Allemandi, Borowiec, \&
  Francaviglia}]{Alle04c}
Allemandi, G., Borowiec, A., \& Francaviglia, M. 2004{\natexlab{b}}, Phys.
  Rev., D70, 103503

\bibitem[{Allemandi {et~al.}(2005)Allemandi, Borowiec, Francaviglia, \&
  Odintsov}]{Alle05}
Allemandi, G., Borowiec, A., Francaviglia, M., \& Odintsov, S.~D. 2005, Phys.
  Rev., D72, 063505

\bibitem[{Allemandi {et~al.}(2006)Allemandi, Capone, Capozziello, \&
  Francaviglia}]{Alle04}
Allemandi, G., Capone, M., Capozziello, S., \& Francaviglia, M. 2006, Gen. Rel.
  Grav., 38, 33

\bibitem[{Amarzguioui {et~al.}(2005)Amarzguioui, Elgaroy, \&
  Multamaki}]{Amar04}
Amarzguioui, M., Elgaroy, O., \& Multamaki, T. 2005, JCAP, 0501, 008

\bibitem[{Amendola(2000)}]{Amen99}
Amendola, L. 2000, Phys. Rev., D62, 043511

\bibitem[{Arkani-Hamed {et~al.}(2002)Arkani-Hamed, Dimopoulos, Dvali, \&
  Gabadadze}]{Arka02}
Arkani-Hamed, N., Dimopoulos, S., Dvali, G., \& Gabadadze, G. 2002,
  hep-th/0209227

\bibitem[{Barraco {et~al.}(2002)Barraco, Hamity, \& Vucetich}]{Barr02}
Barraco, D., Hamity, V.~H., \& Vucetich, H. 2002, Gen. Rel. Grav., 34, 533

\bibitem[{Barraco {et~al.}(2000)Barraco, Hamity, \& F.}]{Barr00}
Barraco, D.~E., Hamity, V.~H., \& F., F. M.~A. 2000, Phys. Rev., D62, 044027

\bibitem[{Barrow \& Clifton(2006)}]{Barr05}
Barrow, J.~D. \& Clifton, T. 2006, Class. Quant. Grav., 23, L1

\bibitem[{Barrow \& Cotsakis(1988)}]{Barr88}
Barrow, J.~D. \& Cotsakis, S. 1988, Phys. Lett., B214, 515

\bibitem[{Barrow \& Cotsakis(1991)}]{Barr91}
Barrow, J.~D. \& Cotsakis, S. 1991, Phys. Lett., B258, 299

\bibitem[{Barrow \& Ottewill(1983)}]{Barr83}
Barrow, J.~D. \& Ottewill, A.~C. 1983, J. Phys., A16, 2757

\bibitem[{Bekenstein(2005)}]{mond3}
Bekenstein, J.~D. 2005, PoS, JHW2004, 012

\bibitem[{Bento {et~al.}(2002)Bento, Bertolami, \& Sen}]{Bent02a}
Bento, M.~C., Bertolami, O., \& Sen, A.~A. 2002, Phys. Rev., D66, 043507

\bibitem[{Bento {et~al.}(2003)Bento, Bertolami, \& Sen}]{Bent02b}
Bento, M.~C., Bertolami, O., \& Sen, A.~A. 2003, Phys. Rev., D67, 063003

\bibitem[{Bertone {et~al.}(2005)Bertone, Hooper, \& Silk}]{Bert04}
Bertone, G., Hooper, D., \& Silk, J. 2005, Phys. Rept., 405, 279

\bibitem[{Birrell \& Davies(1982)}]{Birr82}
Birrell, N.~D. \& Davies, P. 1982, Quantum fields in curved space (Cambridge:
  Cambridge University Press)

\bibitem[{Bond {et~al.}(1997)Bond, Efstathiou, \& Tegmark}]{Bond97}
Bond, J.~R., Efstathiou, G., \& Tegmark, M. 1997, Mon. Not. Roy. Astron. Soc.,
  291, L33

\bibitem[{Brodbeck \& Straumann(1993)}]{Brod92}
Brodbeck, O. \& Straumann, N. 1993, J. Math. Phys., 34, 2412

\bibitem[{Bronnikov \& Chernakova(2005)}]{Bron05}
Bronnikov, K.~A. \& Chernakova, M.~S. 2005, gr-qc/0503025

\bibitem[{Bronnikov \& Melnikov(1995)}]{Bron94}
Bronnikov, K.~A. \& Melnikov, V.~N. 1995, Gen. Rel. Grav., 27, 465

\bibitem[{Brookfield {et~al.}(2005)Brookfield, van~de Bruck, Mota, \&
  Tocchini-Valentini}]{Brook05}
Brookfield, A.~W., van~de Bruck, C., Mota, D.~F., \& Tocchini-Valentini, D.
  2005, astro-ph/0503349

\bibitem[{Buchbinder {et~al.}(1992)Buchbinder, Odintsov, \& Shapiro}]{Buch92}
Buchbinder, I.~L., Odintsov, S.~D., \& Shapiro, I.~L. 1992, Effective action in
  quantum gravity (Bristol: Institute of Physics Publ.)

\bibitem[{Capozziello(2002)}]{Capo02}
Capozziello, S. 2002, Int. J. Mod. Phys., D11, 483

\bibitem[{Capozziello {et~al.}(2005)Capozziello, Cardone, \& Troisi}]{Capo05}
Capozziello, S., Cardone, V.~F., \& Troisi, A. 2005, Phys. Rev., D71, 043503

\bibitem[{Carloni {et~al.}(2005)Carloni, Dunsby, Capozziello, \&
  Troisi}]{Carl04}
Carloni, S., Dunsby, P. K.~S., Capozziello, S., \& Troisi, A. 2005, Class.
  Quant. Grav., 22, 4839

\bibitem[{Carroll {et~al.}(2005)Carroll, De~Felice, Duvvuri, Easson, Trodden,
  \& Turner}]{Carr04}
Carroll, S.~M., De~Felice, A., Duvvuri, V., {et~al.} 2005, Phys. Rev., D71,
  063513

\bibitem[{Cembranos(2005)}]{Cemb05}
Cembranos, J. A.~R. 2005, gr-qc/0507039

\bibitem[{Chiba(2003)}]{Chib03}
Chiba, T. 2003, Phys. Lett., B575, 1

\bibitem[{Clifton \& Barrow(2005)}]{Clif05}
Clifton, T. \& Barrow, J.~D. 2005, Phys. Rev., D72, 103005

\bibitem[{Clifton {et~al.}(2005)Clifton, Mota, \& Barrow}]{Clif04b}
Clifton, T., Mota, D.~F., \& Barrow, J.~D. 2005, Mon. Not. Roy. Astron. Soc.,
  358, 601

\bibitem[{Cognola {et~al.}(2005)Cognola, Elizalde, Nojiri, Odintsov, \&
  Zerbini}]{Cogn05}
Cognola, G., Elizalde, E., Nojiri, S., Odintsov, S.~D., \& Zerbini, S. 2005,
  JCAP, 0502, 010

\bibitem[{Colless {et~al.}(2001)}]{2dFGRS}
Colless, M. {et~al.} 2001, Mon. Not. Roy. Astron. Soc., 328, 1039

\bibitem[{Deffayet(2001)}]{Deff00}
Deffayet, C. 2001, Phys. Lett., B502, 199

\bibitem[{Deffayet {et~al.}(2002)Deffayet, Landau, Raux, Zaldarriaga, \&
  Astier}]{Deff02}
Deffayet, C., Landau, S.~J., Raux, J., Zaldarriaga, M., \& Astier, P. 2002,
  Phys. Rev., D66, 024019

\bibitem[{Dolgov \& Kawasaki(2003)}]{Dolg03}
Dolgov, A.~D. \& Kawasaki, M. 2003, Phys. Lett., B573, 1

\bibitem[{Dominguez \& Barraco(2004)}]{Domi04}
Dominguez, A.~E. \& Barraco, D.~E. 2004, Phys. Rev., D70, 043505

\bibitem[{Dvali \& Turner(2003)}]{Dval03}
Dvali, G. \& Turner, M.~S. 2003, astro-ph/0301510

\bibitem[{Dvali {et~al.}(2000)Dvali, Gabadadze, \& Porrati}]{Dval02}
Dvali, G.~R., Gabadadze, G., \& Porrati, M. 2000, Phys. Lett., B485, 208

\bibitem[{Eidelman {et~al.}(2004)}]{particlereview}
Eidelman, S. {et~al.} 2004, Phys. Lett., B592, 1

\bibitem[{Eisenstein \& Hu(1998)}]{Eise97}
Eisenstein, D.~J. \& Hu, W. 1998, Astrophys. J., 496, 605

\bibitem[{Eisenstein \& White(2004)}]{Eise04}
Eisenstein, D.~J. \& White, M.~J. 2004, Phys. Rev., D70, 103523

\bibitem[{Eisenstein {et~al.}(2005)}]{Eise05}
Eisenstein, D.~J. {et~al.} 2005, Astrophys. J., 633, 560

\bibitem[{Elgaroy \& Multamaki(2005)}]{Elga04}
Elgaroy, O. \& Multamaki, T. 2005, Mon. Not. Roy. Astron. Soc., 356, 475

\bibitem[{Ellis(2000)}]{Ellis98}
Ellis, J.~R. 2000, Phys. Scripta, T85, 221

\bibitem[{Ezawa {et~al.}(2003)Ezawa, Iwasaki, Ohkuwa, Yamada, \& Yano}]{Ezaw03}
Ezawa, Y., Iwasaki, H., Ohkuwa, Y., Yamada, N., \& Yano, T. 2003, gr-qc/0309010

\bibitem[{Flanagan(2004)}]{Flan03}
Flanagan, E.~E. 2004, Phys. Rev. Lett., 92, 071101

\bibitem[{Freese \& Lewis(2002)}]{Free02}
Freese, K. \& Lewis, M. 2002, Phys. Lett., B540, 1

\bibitem[{Gasperini \& Veneziano(1992)}]{Gasp91}
Gasperini, M. \& Veneziano, G. 1992, Phys. Lett., B277, 256

\bibitem[{Gong {et~al.}(2004)Gong, Chen, \& Duan}]{Gong04}
Gong, Y.-G., Chen, X.-M., \& Duan, C.-K. 2004, Mod. Phys. Lett., A19, 1933

\bibitem[{Hawkins {et~al.}(2003)}]{Hawk02}
Hawkins, E. {et~al.} 2003, Mon. Not. Roy. Astron. Soc., 346, 78

\bibitem[{Hu \& Sugiyama(1996)}]{Hu95}
Hu, W. \& Sugiyama, N. 1996, Astrophys. J., 471, 542

\bibitem[{Hwang(1991{\natexlab{a}})}]{hwan91a}
Hwang, J.~C. 1991{\natexlab{a}}, Class. Quant. Grav., 8, 195

\bibitem[{Hwang(1991{\natexlab{b}})}]{hwan91b}
Hwang, J.-C. 1991{\natexlab{b}}, Astrophys. J., 375, 443

\bibitem[{Hwang \& Noh(2001)}]{Hwan01}
Hwang, J.-c. \& Noh, H. 2001, Phys. Lett., B506, 13

\bibitem[{Knop {et~al.}(2003)}]{Knop03}
Knop, R.~A. {et~al.} 2003, Astrophys. J., 598, 102

\bibitem[{Koivisto(2005)}]{Koiv05a}
Koivisto, T. 2005, gr-qc/0505128

\bibitem[{Koivisto \& Kurki-Suonio(2005)}]{Koiv05b}
Koivisto, T. \& Kurki-Suonio, H. 2005, astro-ph/0509422

\bibitem[{Koivisto {et~al.}(2005)Koivisto, Kurki-Suonio, \& Ravndal}]{Koiv04}
Koivisto, T., Kurki-Suonio, H., \& Ravndal, F. 2005, Phys. Rev., D71, 064027

\bibitem[{La \& Steinhardt(1989)}]{La89}
La, D. \& Steinhardt, P.~J. 1989, Phys. Rev. Lett., 62, 376

\bibitem[{Lima(2004)}]{Lima04}
Lima, J.~A.~S. 2004, Braz. J. Phys., 34, 194

\bibitem[{Linder(2003)}]{Lind03}
Linder, E.~V. 2003, Phys. Rev., D68, 083504

\bibitem[{Linder(2005)}]{Lind05}
Linder, E.~V. 2005, astro-ph/0507308

\bibitem[{Lue(2003)}]{Lue02}
Lue, A. 2003, Phys. Rev., D67, 064004

\bibitem[{Lue {et~al.}(2004{\natexlab{a}})Lue, Scoccimarro, \&
  Starkman}]{Lue03}
Lue, A., Scoccimarro, R., \& Starkman, G.~D. 2004{\natexlab{a}}, Phys. Rev.,
  D69, 044005

\bibitem[{Lue {et~al.}(2004{\natexlab{b}})Lue, Scoccimarro, \&
  Starkman}]{Lue04}
Lue, A., Scoccimarro, R., \& Starkman, G.~D. 2004{\natexlab{b}}, Phys. Rev.,
  D69, 124015

\bibitem[{Magnano \& Sokolowski(1994)}]{Magn93}
Magnano, G. \& Sokolowski, L.~M. 1994, Phys. Rev., D50, 5039

\bibitem[{Melchiorri {et~al.}(2003)Melchiorri, Mersini-Houghton, Odman, \&
  Trodden}]{Melc02}
Melchiorri, A., Mersini-Houghton, L., Odman, C.~J., \& Trodden, M. 2003, Phys.
  Rev., D68, 043509

\bibitem[{Mena {et~al.}(2005)Mena, Santiago, \& Weller}]{Mena05}
Mena, O., Santiago, J., \& Weller, J. 2005, astro-ph/0510453

\bibitem[{Meng \& Wang(2003)}]{Meng03}
Meng, X. \& Wang, P. 2003, Class. Quant. Grav., 20, 4949

\bibitem[{Meng \& Wang(2004)}]{Meng03b}
Meng, X.-H. \& Wang, P. 2004, Phys. Lett., B584, 1

\bibitem[{Milgrom(1994)}]{mond1}
Milgrom, M. 1994, Ann. Phys., 229, 384

\bibitem[{Moffat(2004)}]{mond2}
Moffat, J.~W. 2004, astro-ph/0403266

\bibitem[{Mota \& Barrow(2004{\natexlab{a}})}]{Mota03a}
Mota, D.~F. \& Barrow, J.~D. 2004{\natexlab{a}}, Mon. Not. Roy. Astron. Soc.,
  349, 281

\bibitem[{Mota \& Barrow(2004{\natexlab{b}})}]{Mota03b}
Mota, D.~F. \& Barrow, J.~D. 2004{\natexlab{b}}, Phys. Lett., B581, 141

\bibitem[{Mota \& van~de Bruck(2004)}]{Mota04}
Mota, D.~F. \& van~de Bruck, C. 2004, Astron. Astrophys., 421, 71

\bibitem[{Multamaki \& Elgaroy(2004)}]{Mult03c}
Multamaki, T. \& Elgaroy, O. 2004, Astron. Astrophys., 423, 811

\bibitem[{Multamaki {et~al.}(2003)Multamaki, Gaztanaga, \& Manera}]{Mult03a}
Multamaki, T., Gaztanaga, E., \& Manera, M. 2003, Mon. Not. Roy. Astron. Soc.,
  344, 761

\bibitem[{Multamaki {et~al.}(2004)Multamaki, Manera, \& Gaztanaga}]{Mult03b}
Multamaki, T., Manera, M., \& Gaztanaga, E. 2004, Phys. Rev., D69, 023004

\bibitem[{Neville(1980)}]{Nevi79}
Neville, D.~E. 1980, Phys. Rev., D21, 2770

\bibitem[{Nojiri \& Odintsov(2003)}]{Noji03b}
Nojiri, S. \& Odintsov, S.~D. 2003, Phys. Rev., D68, 123512

\bibitem[{Nojiri \& Odintsov(2004{\natexlab{a}})}]{Noji04}
Nojiri, S. \& Odintsov, S.~D. 2004{\natexlab{a}}, PoS, WC2004, 024

\bibitem[{Nojiri \& Odintsov(2004{\natexlab{b}})}]{Noji03a}
Nojiri, S. \& Odintsov, S.~D. 2004{\natexlab{b}}, Gen. Rel. Grav., 36, 1765

\bibitem[{Nojiri \& Odintsov(2005)}]{Noji05}
Nojiri, S. \& Odintsov, S.~D. 2005, Phys. Lett., B631, 1

\bibitem[{Nunez \& Solganik(2004)}]{Nune04}
Nunez, A. \& Solganik, S. 2004, hep-th/0403159

\bibitem[{Odman {et~al.}(2003)Odman, Melchiorri, Hobson, \& Lasenby}]{Odma02}
Odman, C.~J., Melchiorri, A., Hobson, M.~P., \& Lasenby, A.~N. 2003, Phys.
  Rev., D67, 083511

\bibitem[{Olmo(2005{\natexlab{a}})}]{Olmo05a}
Olmo, G.~J. 2005{\natexlab{a}}, gr-qc/0505135

\bibitem[{Olmo(2005{\natexlab{b}})}]{Olmo05b}
Olmo, G.~J. 2005{\natexlab{b}}, gr-qc/0505136

\bibitem[{Padmanabhan(1993)}]{Padm93}
Padmanabhan, T. 1993, Structure Formation in the Universe (Cambridge: Cambridge
  University Press)

\bibitem[{Peebles(1993)}]{Peeb93}
Peebles, P.~J.~E. 1993, Principles of physical cosmology (Princeton, N.J.:
  Princeton University Press)

\bibitem[{Quandt \& Schmidt(1991)}]{Quan90}
Quandt, I. \& Schmidt, H.-J. 1991, Astron. Nachr., 312, 97

\bibitem[{Ramaswamy \& Yasskin(1979)}]{Rama79}
Ramaswamy, S. \& Yasskin, P.~B. 1979, Phys. Rev., D19, 2264

\bibitem[{Riess {et~al.}(2004)}]{Riess}
Riess, A.~G. {et~al.} 2004, Astrophys. J., 607, 665

\bibitem[{Rippl {et~al.}(1996)Rippl, van Elst, Tavakol, \& Taylor}]{Ripp95}
Rippl, S., van Elst, H., Tavakol, R.~K., \& Taylor, D. 1996, Gen. Rel. Grav.,
  28, 193

\bibitem[{Sanders \& McGaugh(2002)}]{mond}
Sanders, R.~H. \& McGaugh, S.~S. 2002, Ann. Rev. Astron. Astrophys., 40, 263

\bibitem[{Schmidt(1998)}]{Schm98}
Schmidt, H.~J. 1998, gr-qc/9808060

\bibitem[{Sellwood \& Kosowsky(2001)}]{mond5}
Sellwood, J.~A. \& Kosowsky, A. 2001, astro-ph/0109555

\bibitem[{Shao {et~al.}(2005)Shao, Gui, \& Wang}]{Shao05}
Shao, Y., Gui, Y.-X., \& Wang, W. 2005, astro-ph/0509158

\bibitem[{Skordis {et~al.}(2006)Skordis, Mota, Ferreira, \& Boehm}]{mond4}
Skordis, C., Mota, D.~F., Ferreira, P.~G., \& Boehm, C. 2006, Phys. Rev. Lett.,
  96, 011301

\bibitem[{Sotiriou(2005{\natexlab{a}})}]{Soti05a}
Sotiriou, T.~P. 2005{\natexlab{a}}, gr-qc/0507027

\bibitem[{Sotiriou(2005{\natexlab{b}})}]{Soti05b}
Sotiriou, T.~P. 2005{\natexlab{b}}, gr-qc/0509029

\bibitem[{Spergel {et~al.}(2003)}]{wmap}
Spergel, D.~N. {et~al.} 2003, Astrophys. J. Suppl., 148, 175

\bibitem[{Starobinsky(1980)}]{Star80}
Starobinsky, A.~A. 1980, Phys. Lett., B91, 99

\bibitem[{Tegmark {et~al.}(2004)}]{sloan}
Tegmark, M. {et~al.} 2004, Astrophys. J., 606, 702

\bibitem[{Verde {et~al.}(2002)}]{Verd01}
Verde, L. {et~al.} 2002, Mon. Not. Roy. Astron. Soc., 335, 432

\bibitem[{Vilkovisky(1992)}]{Vilk92}
Vilkovisky, G.~A. 1992, Class. Quant. Grav., 9, 895

\bibitem[{Vollick(2003)}]{Voll03b}
Vollick, D.~N. 2003, Phys. Rev., D68, 063510

\bibitem[{Vollick(2004)}]{Voll03}
Vollick, D.~N. 2004, Class. Quant. Grav., 21, 3813

\bibitem[{Wang \& Meng(2004)}]{Wang04b}
Wang, P. \& Meng, X.-H. 2004, TSPU Vestnik, 44N7, 40

\bibitem[{Wang \& Tegmark(2004)}]{Wang04}
Wang, Y. \& Tegmark, M. 2004, Phys. Rev. Lett., 92, 241302

\end{thebibliography}

\end{document}